\documentclass[aps,prb,twocolumn,superscriptaddress,floatfix]{revtex4-2}
\usepackage{graphicx}
\usepackage[hidelinks]{hyperref}
\usepackage{dsfont}
\usepackage{xspace}
\usepackage{makecell}
\usepackage{threeparttable}

\usepackage{booktabs}
\usepackage{siunitx}
\DeclareSIUnit\angstromunit{\text{\AA}}
\usepackage{multirow}
\usepackage{amsmath}
\usepackage[final]{changes}

\newcommand{\wj}[1]{{\color[rgb]{.2,.1,.7}{#1}}}

\def\muB{$\mu_{\rm B}$\xspace}
\def\Erion{Er$^{3+}$\xspace}
\def\NBEB{NaBaEr(BO$_3$)$_2$\xspace}

\begin{document}
	
	
    %
    \title{Contrasting Crystal-Field and Dipolar Anisotropies in the Triangular Antiferromagnet \NBEB}
	
	
	
	\author{Cheng Su}
	\affiliation{School of Physics, Beihang University, Beijing 100191, China}

	\author{Yonglin Wang}
	\affiliation{School of Physics, Beihang University, Beijing 100191, China}
	
	\author{Zihang Gao}
	\affiliation{School of Physics, Beihang University, Beijing 100191, China}

    \author{Xinyang Liu}
    \affiliation{School of Physics, Beihang University, Beijing 100191, China}
    \affiliation{Beijing National Laboratory for Condensed Matter Physics, Institute of Physics, Chinese Academy of Sciences, 100190 Beijing, China}

    \author{Jianglong Zhang}
    \affiliation{School of Physical Science and Technology, ShanghaiTech University, Shanghai 201210, China}
    \affiliation{ShanghaiTech Laboratory for Topological Physics, ShanghaiTech University, Shanghai 201210, China}

	\author{Chin-Wei Wang}
	\affiliation{National Synchrotron Radiation Research Center, Hsinchu 30077, Taiwan}
	
	\author{Michael Marek Koza}
	\affiliation{Institut Laue-Langevin, 71 avenue des Martyrs CS 20156, Grenoble, Cedex 9, France}

    \author{Juping Xu}
	\affiliation{Spallation Neutron Source Science Center, Dongguan 523803, China}
    
	\author{Wen Yin}
	\affiliation{Spallation Neutron Source Science Center, Dongguan 523803, China}

    \author{Hao Deng}
    \affiliation{School of Physical Science and Technology, ShanghaiTech University, Shanghai 201210, China}
    \affiliation{ShanghaiTech Laboratory for Topological Physics, ShanghaiTech University, Shanghai 201210, China}

    \author{Wei Li}
    \affiliation{CAS Key Laboratory of Theoretical Physics, Institute of Theoretical Physics, Chinese Academy of Sciences, Beijing 100190, China}
    \affiliation{Peng Huanwu Collaborative Center for Research and Education, Beihang University, Beijing 100191, China}
	
	\author{Junsen Xiang}
	\affiliation{Beijing National Laboratory for Condensed Matter Physics, Institute of Physics, Chinese Academy of Sciences, 100190 Beijing, China}
	
	\author{Wentao Jin}
	\email{wtjin@buaa.edu.cn}
	\affiliation{School of Physics, Beihang University, Beijing 100191, China}

	
	\date{\today}
	
	\begin{abstract}
	The magnetic ground states of rare-earth frustrated magnets can be governed by competing energy scales arising from crystal-field anisotropy, exchange interactions, and long-range dipolar interactions.
    Here we investigate this interplay in the triangular-lattice antiferromagnet \NBEB ~by combining magnetization, specific heat, and neutron scattering measurements.
    Neutron powder diffraction resolves rotations of the ErO$_6$ octahedra and supports the lower-symmetry $R\bar{3}$ structure over the previously proposed $R\bar{3}m$ model.
    Magnetization measurements and inelastic neutron scattering establish an effective spin-1/2 description of its low-temperature magnetism with pronounced easy-axis anisotropy, arising from a well-isolated Kramers ground-state doublet separated from the first excited crystal-electric-field level by approximately 4~meV.
    Below $T_{\rm N}=0.1$~K, single-crystal neutron diffraction reveals A-type antiferromagnetic order with the ordered moment predominantly oriented within the $ab$ plane, despite the pronounced paramagnetic easy-axis anisotropy.
    The estimated dipolar interaction scale is comparable to $T_{\rm N}$, and classical dipolar-energy calculations for the observed stacking pattern favor an in-plane moment orientation.
    These results reveal contrasting crystal-field and dipolar anisotropies in determining the magnetic ground state of \NBEB.
    \end{abstract}

	\maketitle
	
    
	\section{Introduction}
    
    
    Frustrated triangular-lattice antiferromagnets (TLAFs) provide a fertile platform for exploring magnetic ground states arising from competing interactions.
    In an ideal TLAF, geometrical frustration hinders the simultaneous minimization of antiferromagnetic interactions, leading to significant quantum fluctuations \cite{Wannier1950,Li2022_HeisenbergTLAF,White2007_HeisenbergTLAF}.
    Additional interactions, including further-neighbor exchange and magnetic anisotropy, 
    can further enhance the competition and stabilize a variety of magnetic ground states including quantum spin liquid state, spin-supersolid state, and various field-induced phases \cite{Yamamoto2014,White2015,Sellmann2015,Hu2015_J1J2,Li2016_PhaseDiagram,Maksimov2019,Gao2022,Xiang2024,Gao2024_PRB,Gallegos2025,Shu2026}.

    Rare-earth TLAFs
    can realize a broad range of magnetic anisotropies.
    Strong spin-orbit coupling and the crystal electric field (CEF) split the ionic multiplet
    and lead to effective pseudospin degrees of freedom with direction-dependent magnetic responses at low temperatures, allowing geometrical frustration
    to be studied together with intrinsic magnetic anisotropy.
    This interplay gives rise to a variety of unconventional magnetic ground states and phenomena,
    including quantum-spin-liquid-like behavior in the extensively studied candidate YbMgGaO$_4$ \cite{Shen2016,Paddison2017}, 
    a transverse Ising phase in the non-Kramers system TmMgGaO$_4$ \cite{Hu2020,Li2020,Dun2021},
    and dispersive multipolar excitations in NaErSe$_2$ \cite{Zhang2025}.
    In addition, the large magnetic moments of many rare-earth ions can enhance 
    long-range dipolar interactions, introducing an additional energy scale that can compete with CEF effects and exchange interactions in determining the magnetic ground state and related quantum phase transitions \cite{Kraemer2012,Babkevich2016,Bag2021}.

    The $A$BaRE(BO$_3$)$_2$ ($A$ = Na, K, Rb; RE = rare earth) family has recently emerged
    as a material platform for studying rare-earth triangular-lattice magnetism \cite{Sanders2017,Guo2019NBYB,Guo2019RBRB,Guo2019NBRB,Guo2019KBRB,Pan2021,Tokiwa2021,Jesche2023,Song2023,Shanta2025}.
    This family was initially investigated as luminescent materials \cite{Seryotkin2010,Gao2011,Han2013,Svetlyakova2013,Lian2014,Camardello2015,Kononova2016,Kononova2017} 
    and recently recognized as frustrated magnets hosting well-separated triangular-lattice layers of magnetic ions and sizable magnetocaloric effects (MCEs) \cite{Sanders2017,Guo2019NBYB,Guo2019RBRB,Guo2019NBRB,Guo2019KBRB,Tokiwa2021,Jesche2023,Song2023,Shanta2025}.
    The ability to accommodate a wide range of rare-earth ions, 
    together with the availability of single crystals, 
    enables systematic studies of their magnetic behavior and
    potential applications in sub-Kelvin refrigeration.
    Specifically, in the Na-based compounds, the absence of Ba-Na site mixing
    enables the intrinsic magnetism to be studied free from this source of structural disorder \cite{Guo2019NBYB}.
    Among this family, \NBEB (NBEB) is particularly interesting
    because the \Erion Kramers doublet exhibits strong $g$-factor anisotropy and
    sizable dipolar interactions arising from its large magnetic moment.
    An earlier study reported the crystal growth and bulk magnetic properties of this compound,
    with no thermodynamic transition detected down to 0.35~K \cite{Guo2019NBRB}.
    However, its low-temperature magnetic properties and ground state remain largely unexplored.

    In this work, we investigate the low-temperature magnetism of NBEB through a combination of thermodynamic and neutron scattering measurements.
    Our results reveal a striking contrast between CEF and dipolar anisotropies: although the Er$^{3+}$ ground-state doublet strongly favors the $c$ axis, the ordered moments lie predominantly in the $ab$ plane.
    We show that long-range dipolar interactions, which are non-negligible on the exceptionally small magnetic energy scale of this material, can account for this counterintuitive moment orientation.
    NBEB, therefore, provides an example of how dipolar interactions can compete with CEF anisotropy in governing the magnetic ground state of a rare-earth TLAF.


	\section{Methods}
	Polycrystalline samples of NBEB were synthesized by a solid-state reaction using stoichiometric amounts of Na$_2$CO$_3$ (99.99\%), BaCO$_3$ (99.99\%), H$_3$BO$_3$ (99.99\%), and Er$_2$O$_3$ (99.99\%).
    An additional 3\% excess of Na$_2$CO$_3$, BaCO$_3$, and H$_3$BO$_3$ was added to compensate for 
    potential loss during synthesis. The starting materials were thoroughly mixed, pressed into pellets, and calcined at 960~$^\circ$C with intermediate regrindings to minimize impurity phases. 
	Single-crystal samples of NBEB were grown by the flux method using an H$_3$BO$_3$-NaF flux \cite{Guo2019NBRB}.
    Polycrystalline NBEB and the flux were loaded into a platinum crucible, heated to 960~$^\circ$C, followed by a slow cooling to 790~$^\circ$C at a rate of 2~$^\circ$C/h.
    Pink crystals were obtained after dissolving the residual flux in water.
	For neutron scattering experiments, 95\% $^{11}$B-enriched H$_3$BO$_3$ was used to reduce neutron absorption caused by $^{10}$B.
	
	DC magnetic susceptibility and magnetization measurements were performed using a Quantum Design Magnetic Property Measurement System (MPMS).
    Heat capacity measurements were carried out using a Quantum Design Physical Property Measurement System (PPMS) equipped with a dilution refrigerator insert. 
    All these measurements were performed on single-crystal samples of NBEB.

	Neutron powder diffraction (NPD) measurements were performed on the total-scattering time-of-flight (TOF) diffractometer MPI (Multi-Physics Instrument, 31113.02.CSNS.MPI) at the China Spallation Neutron Source (CSNS) \cite{MPIRef}.
	The polycrystalline sample was loaded into a V--Ni alloy can, and diffraction data were collected at room temperature for structural characterization.
    
	Single-crystal neutron diffraction experiment was conducted on the High-Intensity Powder Diffractometer (Wombat) at the Australian Nuclear Science and Technology Organisation (ANSTO) \cite{WombatRef_New}.
	A dilution refrigerator combined with a vertical-field magnet was used for low-temperature magnetic structure study.
    A 79-mg single crystal was mounted on a copper plate with GE varnish and aligned with the $(H, H, L)$ plane as
	the horizontal scattering plane. 
    The magnetic field was applied along the in-plane $[H,-H,0]$ direction.
    An incident neutron wavelength of 2.41~\AA\xspace was used.
    Rietveld refinements were performed using the FULLPROF program \cite{FULLPROF}.
	
	Inelastic neutron scattering (INS) experiments were carried out on the thermal-neutron time-of-flight spectrometer PANTHER at Institut Laue--Langevin (ILL) \cite{PANTHER_REF}.
    Approximately 7~g of polycrystalline sample was wrapped in aluminum foil and mounted in a cryostat.
    Incident neutron energies of $E_i=30$ and 110~meV were selected 
    to resolve the low-energy CEF excitations and access higher-energy CEF levels, respectively.
    Additional measurements at $E_i=171$, 90, 19, 12.5, and 7.5~meV were performed 
    to optimize the energy resolution and search for possible CEF-level splittings.
	The data were reduced using the MANTID software package \cite{MANTID}.
	
	\section{Results}
    
	\subsection{Refined \texorpdfstring{$R\bar{3}$}{R-3} Crystal Structure}

    \begin{figure*}
        \centering
        \includegraphics[width=1\linewidth]{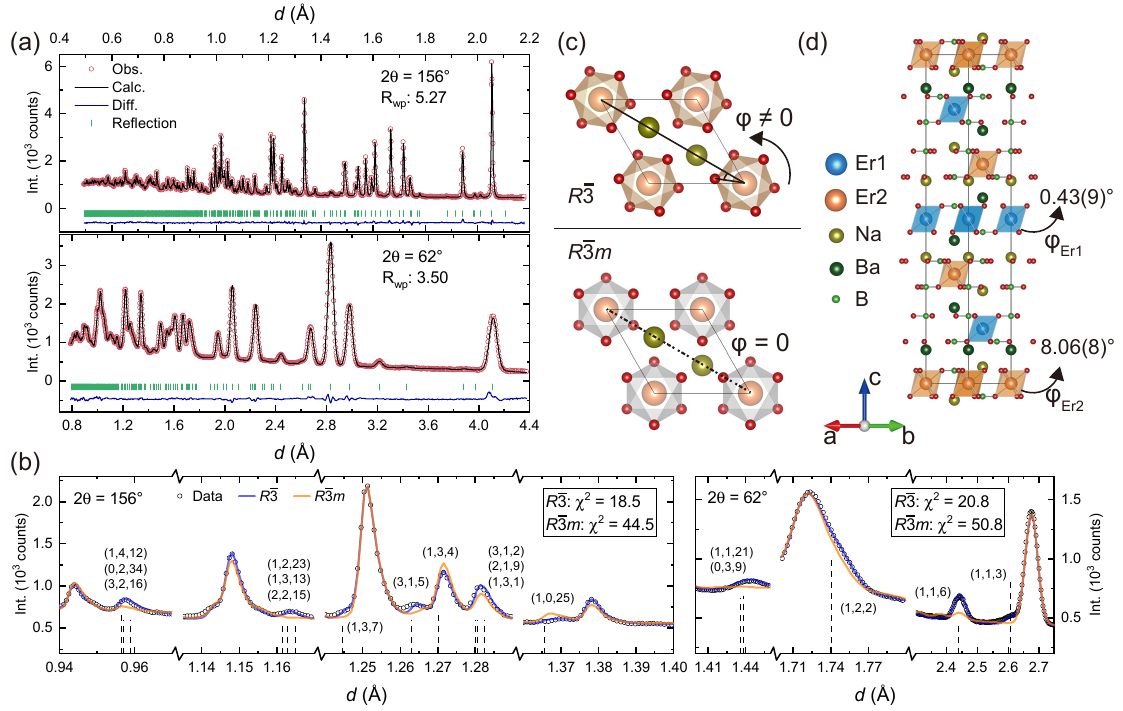}
        \caption{(a) Rietveld refinements of the room-temperature NPD data collected with detector banks located around $2\theta = 156^\circ$ and $62^\circ$.
        (b) Comparison of the refinements in selected regions using the $R\bar{3}$ and $R\bar{3}m$ models.
        The left and right panels correspond to the 156$^\circ$ and 62$^\circ$ detector banks, respectively. The $R\bar{3}$ model provides a significantly better fit to the data.
        (c) Comparison of the $R\bar{3}$ and $R\bar{3}m$ structures viewed along the $c$ axis.
        The angle $\varphi$ denotes the rotation of the ErO$_6$ octahedra about the $c$ axis relative to the mirror-constrained $R\bar{3}m$ configuration with $\varphi=0$.
        (d) Refined crystal structure of NBEB with the $R\bar{3}$ model.
        The different rotation angles $\phi$ for the two crystallographically inequivalent Er sites are indicated.
        }
        \label{fig:Structure}
    \end{figure*}

    NBEB belongs to the layered orthoborate family NaBaRE(BO$_3$)$_2$ (RE = Sc, Y, Tb--Lu) \cite{Seryotkin2010,Svetlyakova2013,Kononova2016}.
    It crystallizes in a layered trigonal structure, 
    in which \Erion occupies the centers of oxygen octahedra and 
    forms triangular layers in the $ab$ plane.
    Along the $c$ axis, these magnetic layers alternate with Na- and Ba-containing borate layers, forming distinct \{Er(Na(BO$_3$))$_2$\}$^-$ and \{Er(Ba(BO$_3$))$_2$\}$^+$ packages in an ABC-stacking sequence.
    Accordingly, \Erion occupies two crystallographically inequivalent sites at the Wyckoff $3a$ (0, 0, 0) and $3b$ (0, 0, 0.5) positions.
    Hereafter, we label the \Erion~ions at the 3$b$ and 3$a$ sites as Er1 and Er2, respectively.

    The crystal structure of NBEB was initially assigned to the space group $R\bar{3}$ 
    based on powder x-ray diffraction (XRD) and 
    its isostructural relation to other NaBaRE(BO$_3$)$_2$ compounds \cite{Kononova2016}.
    A subsequent single-crystal XRD study proposed the higher-symmetry space group $R\bar{3}m$ \cite{Guo2019NBRB}, leaving ambiguity regarding the appropriate structural model.
    The key difference between the two models lies in the symmetry constraints on the oxygen positions.
    In the $R\bar{3}m$ model, mirror symmetry constrains the oxygen atoms to mirror planes, while these constraints are absent in the $R\bar{3}$ model.
    To distinguish between these structural models, we 
    performed neutron powder diffraction, which provides enhanced sensitivity to the oxygen positions.

\begin{table*}
	\caption{\label{Table:Structure}
    Refined structural parameters of NBEB in the $R\bar{3}$ space group.
    The $^{11}$B and $^{10}$B isotopes were constrained to occupy the same crystallographic sites and are therefore not distinguished in the table.
    Their occupancies were fixed according to the nominal isotopic composition of the starting material.}
	\begin{ruledtabular}
		\resizebox{\textwidth}{!}{
		\begin{tabular}{ccccccc}
            \multicolumn{7}{c}{$T$ = 300~K, \qquad $a=b=5.34644(6)$~\AA, \qquad $c=35.7552(7)$~\AA} \\
            \midrule
            Atom & site & symmetry & $x$      & $y$         & $z$    & $B_{\rm iso}$ ($\text{\AA}^2$)\\
			\midrule
			Er (1) & 3b & -3 & 0 & 0 & 0.5         & /\\
			Er (2) & 3a & -3 & 0 & 0 & 0         & /\\
			Ba (1) & 6c & 3 & 0         & 0         & 0.1025(2)       & /\\
			Na (1) & 6c & 3 & 0         & 0       & 0.3841(2)      & /\\
			B (1) & 6c & 3 & 0         & 0         & 0.2056(1)         & 0.80(6)\\
            B (2) & 6c & 3 & 0         & 0         & 0.2940(1)         & 0.42(5)\\
            O (1) & 18f & 1 & 0.3734(5)         & 0.2325(4)         & 0.0394(1)         & /\\
            O (2) & 18f & 1 & 0.9611(6)         & 0.4781(5)         & 0.1275(1)         & /\\
            \midrule
            Atom & $\beta_{11}$  & $\beta_{22}$  & $\beta_{33}$  & $\beta_{12}$  & $\beta_{13}$  & $\beta_{23}$  \\
            Er (1) & 0.008(2) & 0.008(2) & 0.00019(3) & 0.0040(7) & 0         & 0\\
            Er (2) & 0.003(2) & 0.003(2) & 0.00009(3) & 0.0014(6) & 0         & 0\\
            Ba (1) & 0.015(2) & 0.015(2) & 0.00017(2) & 0.0073(6) & 0         & 0\\
            Na (1) & 0.022(3) & 0.022(3) & 0.0017(2) & 0.011(2) & 0         & 0\\
            O (1) & 0.0035(9) & 0.027(2) & 0.0002(1) & 0.0039(8) & -0.0007(1)         & -0.0007(1)\\
            O (2) & 0.009(1) & 0.015(1) & 0.00059(2) & 0.004(1) & -0.0008(1)         & 0.0001(2)\\
		\end{tabular}
		}
	\end{ruledtabular}
	
    \end{table*}
    
    Diffraction data collected from detector banks centered at $2\theta=156^\circ$ and $62^\circ$ were refined simultaneously to cover a broad range of $d$ spacings.
    The data are better described 
    by the $R\bar{3}$ model, as shown by the Rietveld refinements in Fig.~\ref{fig:Structure}(a).
    Selected reflections for which the two models predict markedly different intensities are compared in Fig.~\ref{fig:Structure}(b).
    The $R\bar{3}m$ model fails to reproduce the observed intensities of these reflections and yields anomalously large displacement parameters for the oxygen atoms.
    Removing the mirror-plane constraints on the oxygen positions substantially improves the fits, reducing $\chi^2$ from 44.5 to 18.5 and from 50.8 to 20.8 for the 156$^\circ$ and 62$^\circ$ detector banks, respectively [Fig.~\ref{fig:Structure}(b)].
    The refined structural parameters are summarized in Table~\ref{Table:Structure}.
     
    The additional freedom of the oxygen positions 
    in the $R\bar{3}$ model is manifested as rotations of the ErO$_6$ octahedra 
    away from the mirror-constrained configuration of $R\bar{3}m$ [Fig.~\ref{fig:Structure}(c)].  
    To quantify this distortion,
    we define an octahedral rotation angle $\varphi$,
    with $\varphi=0$ corresponding to the $R\bar{3}m$ configuration.
    At room temperature, the ErO$_6$ octahedra centered at the Er1 and Er2 sites 
    exhibit rotation angles of $\varphi_{\rm Er1}=0.43(9)^\circ$ and $\varphi_{\rm Er2}=8.06(8)^\circ$, respectively [Fig.~\ref{fig:Structure}(d)].
    The substantially larger rotation at the Er2 site reflects the different local environments of the two Er sites:
    the ErO$_6$ octahedra adjacent to the Ba-containing layers remain close to the mirror-symmetric configuration, 
    whereas those adjacent to the Na-containing layers exhibit a more pronounced rotation.
    
    A similar layer-dependent structural distortion has been reported in single-crystal XRD studies of the isostructural compounds NaBaSc(BO$_3$)$_2$ and NaBaY(BO$_3$)$_2$, where the BO$_3$ groups in the Na-containing layers exhibit much larger rotations than those in the Ba-containing layers \cite{Seryotkin2010}.
    This behavior was attributed to the more asymmetric local coordination of Na$^+$ compared with Ba$^{2+}$. 
    The refined $R\bar{3}$ structure establishes the local symmetry of the two Er$^{3+}$ sites and therefore provides the structural basis for the CEF analysis discussed below.
    Our results also suggest that careful structural characterization may be warranted for other NaBaRE(BO$_3$)$_2$ family members.

	\subsection{Bulk Magnetic and Thermodynamic Properties}

    \begin{figure}
        \centering
        \includegraphics[width=1\linewidth]{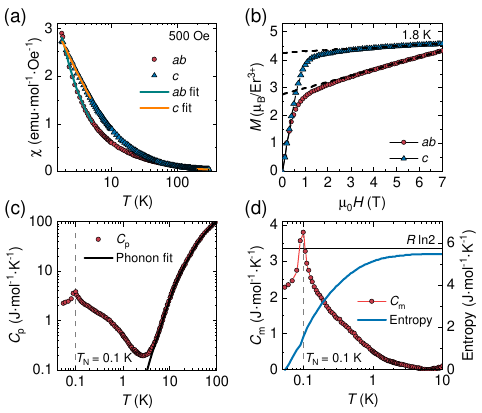}
        \caption{
        (a) DC magnetic susceptibility measured in a field of 500~Oe applied along different directions.
        Solid lines show the CW fits.
        (b) Isothermal magnetization measured at 1.8~K with the field applied along different directions. 
        Dashed lines represent linear fits to the data between 5 and 7~T, used to estimate the Van Vleck contributions.
        (c) Zero-field specific heat measured up to 100~K. 
        The solid line shows the phonon contribution fitted using a three-component Debye model.
        (d) Magnetic specific heat and the corresponding magnetic entropy.
        }
        \label{fig:PhysProp}
    \end{figure}
    
	Fig.~\ref{fig:PhysProp}(a) shows the temperature-dependent dc magnetic susceptibility ($\chi=M/H$) of NBEB.
	The compound remains paramagnetic down to 1.8~K and exhibits a pronounced magnetic anisotropy, with the $c$ direction being the easy axis.
	Curie--Weiss (CW) fits between 200 and 300~K yield effective magnetic moments of 
    $\mu_{\rm eff}=9.58$ and 10.43~\muB for fields applied in the $ab$ plane and along the $c$ axis, respectively,
    close to the free-ion value of 9.59~\muB for \Erion.
	Fits over the low-temperature range of 1.8-5 K yield $\mu_{\rm eff}=6.55$ and 9.25~\muB, 
    with corresponding CW temperatures ($\theta\rm_{CW}$) of $\theta_{\mathrm{CW}}=-0.11$ and $-2.14$ K for $H\parallel ab$ and $H\parallel c$, respectively.
	The negative $\theta\rm_{CW}$ suggests predominantly antiferromagnetic (AFM) correlations, while the pronounced directional dependence reflects the strong magnetic anisotropy.
    
	The isothermal magnetization measured at 1.8~K increases smoothly with applied field [Fig.~\ref{fig:PhysProp}(b)].
	Above approximately 1~T, the magnetization shows an additional nearly linear field dependence, which is attributed to the Van Vleck contribution from excited CEF states.
	Subtracting this linear contribution yields 
    saturated moments of $\mu_c=4.26$~\muB/\Erion and $\mu_{ab}=2.76$~\muB/\Erion, 
    for $H\parallel c$ and $H\parallel ab$, respectively. 
    The larger saturation moment along the $c$ axis further confirms the easy-axis character of the \Erion ground-state doublet.
	
    The zero-field heat capacity of a NBEB single crystal was measured between 0.05 and 100~K.
    A pronounced $\lambda$-like anomaly was observed at $T\rm_N=0.10$~K, 
    signaling a magnetic phase transition [Fig.~\ref{fig:PhysProp}(c,d)].
    This transition temperature is comparable to the magnitude of the low-temperature in-plane $\theta_{\rm CW}$ of $-$0.11 K.
    To estimate the phonon contribution, the data above 6~K were fitted using a three-component Debye model,
    $$
    C\rm_{ph} = 9\it{R}\sum_{i=\rm1}^{\rm3} 
    {c_i (\frac{T}{\theta_{Di}})^{\rm3} 
    \int_{\rm0}^{\theta_{Di}/T}{ \mathrm{d}x  \frac{x^{\rm4} e^x}{(e^x-\rm1)^{2\rm}} } 
    }
    $$
    with the constraint of $c_1 + c_2 + c_3 = 11$, corresponding to the total number of atoms per formula unit.
	The fit yields $c_1$ = 3.42, $c_2$ = 6.83, $c_3$ = 0.75, 
    $\theta_{D1}$ = 240.7~K, $\theta_{D2}$ = 841.2~K and $\theta_{D3}$ = 84.4~K. 
    After subtracting the phonon contribution, the magnetic heat capacity ($C\rm_m$) and corresponding magnetic entropy are obtained, as shown in Fig.~\ref{fig:PhysProp}(d).
    The recovered magnetic entropy approaches $R\ln 2$, consistent with a well-isolated ground-state doublet and an effective spin-1/2 description at low temperatures.

	\begin{figure*}
		\centering
		\includegraphics[width=1\linewidth]{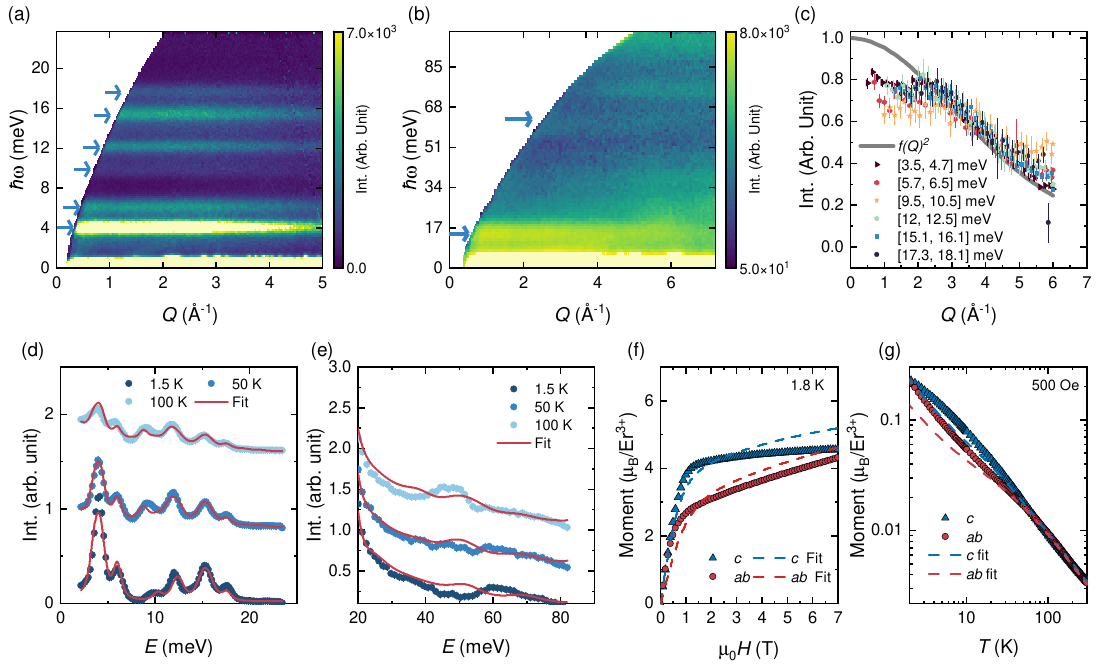}
		\caption{(a,b) Dynamical structure factor $S(Q,\omega)$ measured at 1.5~K with incident energies $E_i=30$~meV (a) and 110~meV (b).
		(c) $Q$ dependence of the integrated intensities of selected excitations. The solid gray line shows the squared \Erion magnetic form factor.
		(d,e) Integrated INS spectra measured with $E_i$ = 30~meV (d) and 110~meV (e) over $Q$ ranges of 1.5--2.5 and 2.7--3.7~\si{\per\angstromunit}, respectively. 
        The solid lines represent the CEF fits.
        Spectra at different temperatures are offset vertically for clarity.
		(f,g) Comparison of the measured magnetic moment as functions of field (f) and temperature (g) with the corresponding CEF fits.
		}
		\label{fig:CEF}
	\end{figure*}
    
	\subsection{CEF Study}
	
	To gain microscopic insight into the pronounced easy-axis anisotropy of NBEB, we performed INS experiments to determine the CEF excitations of \Erion,
	whose $J$ = 15/2 ground-state multiplet is split into eight Kramers doublets at each crystallographically inequivalent site.
    
	At 1.5~K, several dispersionless excitations are observed at 4.0, 6.1, 9.9, 12.1, 15.3, and 17.6~meV [Fig.~\ref{fig:CEF}(a)], with the incident energy of $E_i=30$~meV.
	An additional weak excitation is observed near 63~meV 
    with $E_i=110$~meV [Fig.~\ref{fig:CEF}(b)].
	Due to the limited energy resolution at this incident energy, several excitations resolved in Fig.~\ref{fig:CEF}(a) 
    merge into a broad peak centered near 14~meV in Fig.~\ref{fig:CEF}(b).
	The intensities of the low-energy excitations generally decrease with increasing $Q$ 
    and approximately follow the squared \Erion magnetic form factor [Fig.~\ref{fig:CEF}(c)], supporting their magnetic CEF origin.
	Their intensities also decrease upon warming as the ground-state population is reduced [Figs.~\ref{fig:CEF}(d) and \ref{fig:CEF}(e)].
	The $Q$ dependence of the excitation near 60~meV is less conclusive, but its intensity likewise decreases upon warming.
	At elevated temperatures, an additional excitation emerges near 45~meV [Fig.~\ref{fig:CEF}(e)], 
    consistent with a transition from the thermally populated CEF levels in the 12-18 meV range to the level near 60~meV, further supporting the CEF assignment of the latter. 
	Another excitation near 9~meV develops upon warming [Fig.~\ref{fig:CEF}(d)] and 
    is similarly attributed to a transition between excited CEF states.
    Broad excitations observed at large $Q$ in the $E_i$ = 110~meV spectrum [Fig.~\ref{fig:CEF}(b)] between 10 and 40~meV, and the flat excitation at approximate 75~meV are likely from phonon contributions.
    
	Measurements with lower $E_i$ of 19, 12.5, and 7.5~meV 
    reveal neither additional CEF excitations below 4~meV nor resolvable splitting of the observed modes.
	The first excited CEF level therefore lies at approximately 4~meV, corresponding to an energy scale of about 46~K.
	This energy scale is more than an order of magnitude larger than both $T_{\rm N}$ and the fitted $|\theta_{\rm CW}|$, indicating that excited CEF states are effectively frozen out in the Kelvin regime. 
    Together with the recovery of the magnetic entropy close to $R\ln2$, this supports an effective spin-1/2 description of NBEB's low-temperature magnetism based on a well-isolated Kramers ground-state doublet.

	NBEB contains two crystallographically inequivalent \Erion sites at the Wyckoff $3a$ and $3b$ positions, 
    with distinct local environments associated with the adjacent Na$^+$- and Ba$^{2+}$-containing layers 
    and the different rotations of the ErO$_6$ octahedra (Fig.~\ref{fig:Structure}).
    In the refined $R\bar{3}$ structure, the loss of mirror symmetry reduces
	the local point group of both \Erion sites from  $D_{3d}$ to $S_6$.
	The CEF Hamiltonian can therefore be expressed in terms of Stevens operators as
    \[
    H_{\mathrm{CEF}}
    =
    \sum_{i,n,m}
    B_{n}^{m,(i)} O_{n}^{m,(i)},
    \qquad i=1,2.
    \]
    where $i$ labels the two inequivalent \Erion sites and $(n,m)$ runs over the terms allowed by $S_6$ symmetry.
    Compared with the $R\bar{3}m$ structure, 
    the absence of mirror symmetry permits additional negative-$m$ Stevens terms, 
    highlighting the importance of identifying the correct space group for the CEF analysis.
    In total, $S_6$ symmetry allows nine CEF parameters for each Er site, giving 18 parameters for the two-site model.
    
    A rotation of the local coordinate system about the threefold axis leaves the calculated powder-averaged scattering intensities unchanged, resulting in a redundant degree of freedom in the CEF parameterization.
    This freedom allows one CEF parameter with $m \neq 0$ to be set to zero for each Er site by an appropriate choice of the local coordinate system. 
    We therefore choose the convention
    $$
    B_4^{3,(1)} = B_4^{3,(2)} = 0,
    $$
    reducing the number of independent CEF parameters from 18 to 16.

    The remaining CEF parameters were optimized using the covariance matrix adaptation evolution strategy (CMA--ES) algorithm \cite{CMA-ES,CrysFieldExplorer}, by simultaneously fitting the excitation energies and scattering intensities, as shown in Fig.~\ref{fig:CEF}(d,e).
    The isothermal magnetization and temperature dependence of magnetic moment for fields along the $c$ axis and in the $ab$ plane [Fig.~\ref{fig:CEF}(f,g)] 
    were included as additional constraints.
    The neutron scattering intensities were calculated using an author-modified function based on
    the PyCrystalField package \cite{PyCrystalField}.
    The best-fit results are shown in Fig.~\ref{fig:CEF}(d--g), and the corresponding set of CEF parameters is listed in Table~\ref{Table:CEFparameter}.
    Despite removing the rotational redundancy, the two-site model still
    contains 16 independent Stevens parameters, 
    whereas only a limited number of CEF excitations are resolved experimentally,
    making such a fit underconstrained.
    Therefore, the parameter set in Table~\ref{Table:CEFparameter} should be regarded as a representative solution rather than a unique determination of the CEF Hamiltonian.
    Additional experimental constraints would be required to further restrict the allowed parameter space.
    Nevertheless, the observed excitation spectra together with the bulk magnetic anisotropy support a well-isolated Kramers ground-state doublet with pronounced easy-axis anisotropy.

    	\begin{table}
		\caption{\label{Table:CEFparameter} Representative set of CEF parameters for the two inequivalent \Erion sites. Values are rounded to six significant figures.}
		\begin{ruledtabular}
			\resizebox{\linewidth}{!}{
				\begin{tabular}{ccc}
					Parameters (meV) & site 1 & site 2\\
					\midrule
					$B_2^0$ & $-0.0103389$ & $-0.182133$ \\
					$B_4^{-3}$ & $0.00227568$ & $-0.0244126$\\
					$B_4^0$ & $0.000157387$ & $0.00146296$\\
					$B_4^3$ & 0 & 0\\
					$B_6^{-6}$ & $7.50479\times10^{-5}$ & $-2.86342\times10^{-4}$\\
					$B_6^{-3}$ & $2.57555\times10^{-5}$ & $2.08245\times10^{-4}$\\
					$B_6^0$ & $5.41816\times10^{-6}$ & $5.48289\times10^{-6}$\\
					$B_6^3$ & $-4.05504\times10^{-5}$ & $1.40350\times10^{-4}$\\
					$B_6^6$ & $2.76833\times10^{-7}$ & $-6.32273\times10^{-5}$\\
				\end{tabular}
			}
		\end{ruledtabular}
		
	\end{table}
    
	\subsection{Neutron Diffraction}

	\begin{figure*}
		\centering
		\includegraphics[width=1\linewidth]{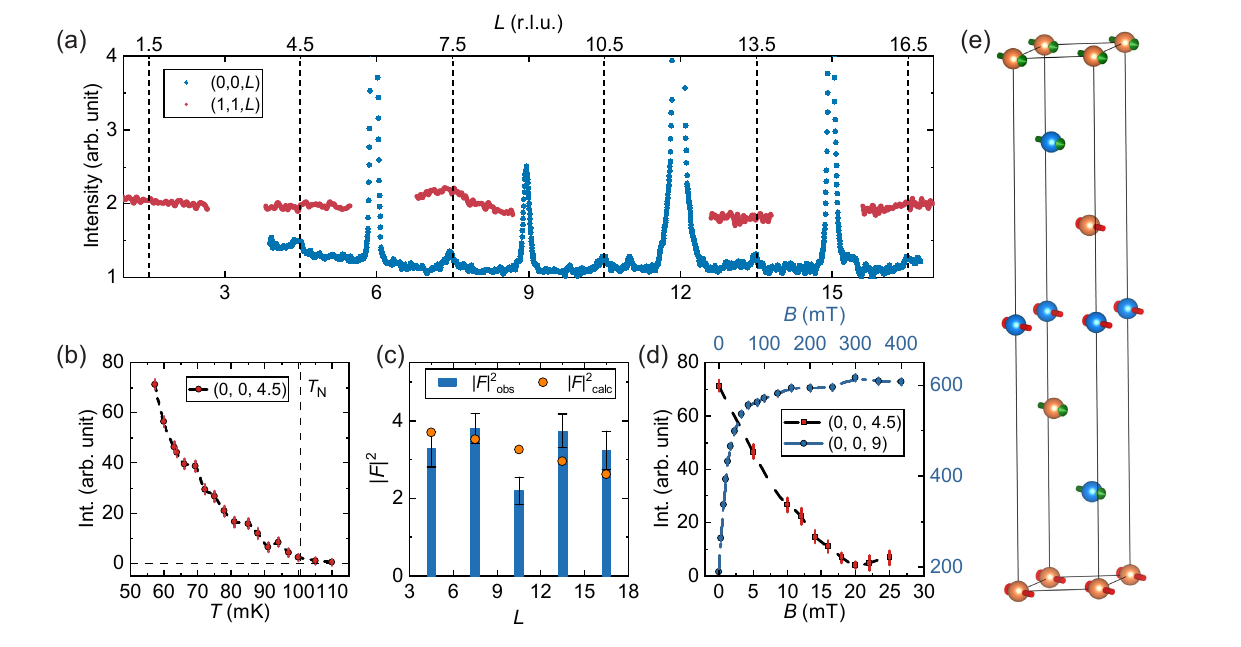}
		\caption{(a) Neutron diffraction intensity scanned along the $(0,0,L)$ and $(1,1,L)$ directions.
        The $(1,1,L)$ data are vertically offset for clarity.
		(b) Temperature dependence of the integrated intensity of the (0, 0, 4.5) magnetic reflection.
		(c) Comparison of the calculated and experimentally observed magnetic structure factors.
		(d) In-plane-field dependence of the integrated intensities of the (0, 0, 4.5) and (0, 0, 9) reflections.
		(e) Minimal Model (Model 2) magnetic structure consistent with the measured $(0,0,L)$ intensities.
        The in-plane moment direction and the precise $c$-axis component cannot be determined. 
        		}
		\label{fig:Wombat}
	\end{figure*}

\begin{table}
	\caption{\label{Table:MagneticStructure}Candidate magnetic-structure models obtained from representation analysis and magnetic space group analysis
    \cite{FULLPROF,Bilbao_1,Bilbao_2,Bilbao_3,Bilbao_ksubgroup_1,Bilbao_ksubgroup_2}.}
	\begin{ruledtabular}
		\resizebox{0.90\linewidth}{!}{
		\begin{threeparttable}
		
			\begin{tabular}{ccccc}
				{Model} & {Irep\tnote{1}} & {MSG\tnote{2}} & {Moment} & {Comment} \\
				\midrule
				\rule{0pt}{0.4cm}
				{1} & {$\Gamma_1$} & {$R_I3$} & {(0,0,$M_c$)} & \makecell[c]{Incompatible with\\$(0,0,L)$ reflections}\\
				\rule{0pt}{0.4cm}
				{2} & {$\Gamma_2 \oplus \Gamma_3 $} & {$P_s1$} & {($M_a$,$M_b$,0)} & {Minimal Model}\\
				\rule{0pt}{0.4cm}
				{3} & {$\Gamma_1 \oplus \Gamma_2 \oplus \Gamma_3$} & {$P_s1$} & {($M_a, M_b, M_c$)} & {Symmetry Allowed}\\
			\end{tabular}
		
			\begin{tablenotes}
				\footnotesize
				\item [1] Irreducible representation
				\item  [2] Magnetic space group
			\end{tablenotes}
		
		\end{threeparttable}
		}
	\end{ruledtabular}
\end{table}

    Single-crystal neutron diffraction was used to determine the magnetic order in NBEB.
    Upon cooling, additional reflections appear at $(0,0,3n+1.5)$ [Fig.~\ref{fig:Wombat}(a)], corresponding to a magnetic propagation vector $\mathbf{k}=(0,0,3/2)$.
    The integrated intensity of the (0, 0, 4.5) reflection decreases continuously upon warming and vanishes near $T_{\rm N}=0.10$~K [Fig.~\ref{fig:Wombat}(b)],
    consistent with the heat-capacity anomaly in Fig.~\ref{fig:PhysProp}(c) and confirming its magnetic origin.
    
    We performed a representation analysis 
    using the BasIReps program implemented in the FULLPROF suite,
    with the space group $R\bar{3}$ and propagation vector $\mathbf{k}=(0,0,3/2)$.
    For each \Erion site, the magnetic representation decomposes as
    $\Gamma_m = \Gamma_1 \oplus \Gamma_2 \oplus \Gamma_3$,
    where $\Gamma_1$ corresponds to moments along the $c$ axis, while $\Gamma_2 \oplus \Gamma_3$ restricts moments within the $ab$ plane (Table~\ref{Table:MagneticStructure}). 
    The magnetic structures considered here have ferromagnetic alignment within each Er layer and AFM stacking between the layers, corresponding to A-type AFM order.
    For this configuration,
    the magnetic structure factor for $(H,H,3n+1.5)$ reflections can be written as
    $$
    \vec{F}(\vec{Q}) = 3 f(Q)[\vec{M}_2 - ie^{-in\pi} \vec{M}_1],
    $$
    where $f(Q)$ is the magnetic form factor of \Erion and $\vec{M}_i$ is the ordered moment on site $i$.
    Because magnetic neutron diffraction probes 
    only the component of $\vec{F}(\vec{Q})$ perpendicular to $\vec{Q}$, 
    the intensities of the observed $(0,0,L)$ magnetic reflections depend only on the in-plane moment components:
    $$
    I(Q) \propto f(Q)^2 (|M_{1}^{ab}|^2 + |M_{2}^{ab}|^2).
    $$
    A structure with moments strictly along the $c$ axis (Model 1) 
    would therefore produce no magnetic intensity for these reflections and can be excluded, despite its higher symmetry retaining the threefold axis, leaving Model~2 ($\Gamma_2 \oplus \Gamma_3$) as the minimal model compatible with the observed reflections.
    Fitting their intensities yields a root-mean-square in-plane ordered moment of 0.9(1)~$\mu_{\rm B}/{\rm Er}^{3+}$ at 57~mK [Fig.~\ref{fig:Wombat}(c)].
    
    In terms of the magnetic-space-group description, 
    Model 2 is the purely in-plane special case of the magnetic space group $P_s1$, 
    which in principle allows a non-zero $c$-axis moment component ($M_c$), as represented by Model 3 [Table~\ref{Table:MagneticStructure}].
    The observed $(0,0,L)$ magnetic reflections therefore 
    establish a finite in-plane component ($M_{ab}$) but are insensitive to the symmetry-allowed $M_c$.
    Nevertheless, the absence of detectable magnetic reflections 
    along the $(1,1,L)$, $(-1,-1,L)$, and $(-2,-2,L)$ directions [Fig.~\ref{fig:Wombat}(a)]  
    places an upper bound on $M_c$.
    With an empirical intensity threshold estimated from our experimental sensitivity and assuming
    the same moment direction for the two \Erion sites,
    we estimate $M_c < 0.35~\mu_{\rm B}$/\Erion for both single-domain and three-domain structures (see Fig.~\ref{figSP:moment-c} in Supplementary Materials for more details).
    These results indicate that the ordered moment lies predominantly in the $ab$ plane,  
    as illustrated in Fig.~\ref{fig:Wombat}(e), 
    despite the easy-axis character of the ground-state Kramers doublet.
    
    The fitted in-plane ordered moment (0.9~$\mu_{\rm B}/{\rm Er}^{3+}$) is 
    substantially smaller than the 
    saturated in-plane moment of 2.76~$\mu_{\rm B}/{\rm Er}^{3+}$ [Fig.~\ref{fig:PhysProp}(b)].
    However, the magnetic intensity has not yet saturated at the lowest measured temperature of 57~mK [Fig.~\ref{fig:Wombat}(b)], suggesting that the ordered moment may increase further upon cooling.
    Measurements at even lower temperatures and over a broader region of reciprocal space would be valuable to better constrain the magnetic structure
    and assess whether quantum fluctuations contribute to the reduced ordered moment \cite{Li2022_HeisenbergTLAF}.
    
    We further examined the field dependence of the magnetic order with an in-plane magnetic field applied along the $(1,-1,0)$ direction.
    As shown in Fig.~\ref{fig:Wombat}(d), the intensity of the $(0,0,4.5)$ reflection is rapidly suppressed with increasing field and vanishes above approximately 20~mT.
    This sensitivity to a very small in-plane field highlights the fragile nature of the A-type AFM order, consistent with the small energy scale reflected by $T_{\rm N}$.
    In contrast, the intensity of the $(0,0,9)$ nuclear reflection increases continuously with field 
    and approaches saturation near 0.1~T, signaling the development of a field-polarized state.

	\section{Discussion and Conclusion}
	
	   	\begin{figure}
		\centering
		\includegraphics[width=1\linewidth]{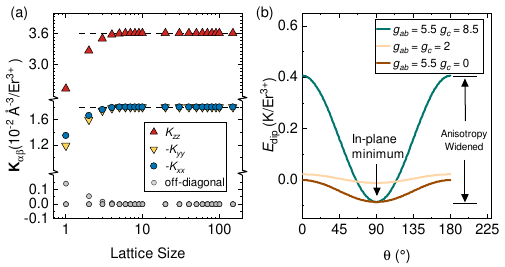}
		\caption{(a) Convergence of the calculated $\mathbf K_{\alpha\beta}$ with increasing lattice size. The diagonal elements converge to finite values, while the off-diagonal terms converge to zero.
		(b) Angular dependence of the calculated dipolar energy ($E\rm_{dip}$) for different $g$-factor anisotropies,
        including $g_{ab} = 5.5$ and $g_c=8.5$ inferred from the magnetization, 
        the fully easy-plane case ($g_c=0$), and the isotropic case ($g_{ab} = g_c = 2$). 
        Here, $\theta$ denotes the angle between the spin direction and the $c$ axis.} 
		\label{fig:DipolarSum}
	\end{figure}

    Our results establish NBEB as an effective spin-1/2 magnet at low temperatures, with a small ordering scale evidenced by $T_{\rm N}=0.10$~K and the suppression of magnetic order by a 20~mT in-plane field.
    Despite the pronounced easy-axis anisotropy of the \Erion ground-state doublet, the ordered moments lie predominantly in the $ab$ plane, suggesting a competition between local magnetic anisotropy and the interactions that select the ordered-moment orientation.
    
    For ions with large magnetic moments, such as \Erion, dipole--dipole interactions can 
    provide an additional source of magnetic anisotropy.
    Their characteristic energy scale in NBEB can be estimated as
    $$
    E_{\rm dip}/{k_{\rm B}} = \frac{\mu_{0} \mu^2}{4\pi k_{\rm B}a^3} \approx 74~{\rm mK},
    $$
    using $\mu=4.26~\mu_{\rm B}$ and $a=5.34644$~\AA\ inferred from the magnetization measurement and structural characterization.
    This energy scale is comparable to $T_{\rm N}$ and the magnitude of the in-plane $|\theta_{\rm CW}|$, suggesting that
    dipolar interactions cannot be neglected in the low-energy magnetic Hamiltonian.
    
    Previous theoretical studies have shown that classical dipoles on an infinite two-dimensional triangular lattice favor a ferromagnetic arrangement when only dipole--dipole interactions are considered \cite{Rastelli2003_DipoleMC,Johnston2016_DipoleCalc},
    which is consistent with the ferromagnetic alignment within each \Erion layer in the A-type AFM order observed in NBEB.
    However, when AFM exchange is also present and comparable to the dipolar coupling, 
    their competition can instead stabilize other magnetic ground states such as stripe order \cite{Rastelli2003_DipoleMC}.
    Such stripe order has been observed in several Er-based TLAFs, 
    including KErSe$_2$ \cite{Ding2023_KErSe2} and NaErSe$_2$ \cite{Zhang2025}, 
    where sizable off-diagonal interactions have been identified by neutron scattering measurements
    and may, in part, originate from the dipolar interactions.
    By contrast, within the dipolar-Heisenberg model \cite{Rastelli2003_DipoleMC}, 
    the intralayer ferromagnetic arrangement observed in our case, together with the comparable dipolar and experimentally inferred magnetic energy scales, 
    suggests that NBEB lies closer to the dipolar-dominated regime with relatively weak exchange interactions.
    
    Motivated by these considerations, we evaluate the classical dipolar energy for the experimentally determined A-type AFM structure with $\mathbf{k}=(0,0,3/2)$.
    The average dipolar energy per \Erion ion is given by
    $$
    E\rm_{dip} = \frac{\mu_0}{4\pi} \frac{1}{2N_m}  \sum_{i,j, \mathbf{R}}  \left[
    \frac{\boldsymbol{\mu}_i\cdot\boldsymbol{\mu}_j}{r_{ij,\mathbf R}^{3}}
     - 3\frac{
    (\boldsymbol{\mu}_i\cdot\mathbf r_{ij,\mathbf R})
    (\boldsymbol{\mu}_j\cdot\mathbf r_{ij,\mathbf R})
    }{r_{ij,\mathbf R}^{5}}
     \right],
    $$
    where $i$ and $j$ label different \Erion sites within the magnetic unit cell, $\mathbf{R}$ labels lattice translations, and $N_m$ is the number of \Erion ions in the magnetic unit cell. 
    For a collinear structure, the magnetic moment at site $i$ is written as $\boldsymbol{\mu}_i = b_i\mu_{\rm B}S\,\mathbf G\,\hat{\mathbf n}$ 
    $(b_i=\pm1) $, where $b_i \hat{\mathbf{n}}$ specifies the spin direction with $\hat{\mathbf n}$ being the unit vector and $\mathbf{G}$ is the $g$-tensor of the \Erion ground-state CEF doublet.
    Assuming the same spin direction and the same $g$-tensor for the two Er$^{3+}$ sites, the dipolar energy can then be reduced to

    $$
    E_{\rm dip}
    =
    \frac{\mu_0\mu_{\rm B}^{2}S^{2}}{4\pi}
    \hat{\mathbf n}^{T}
    \mathbf G^{T}
    \mathbf K
    \mathbf G
    \hat{\mathbf n}.
    $$
    where the geometry-dependent matrix $\mathbf{K}$ decouples the lattice sum from the moment direction:
    $$\mathbf K = \frac{1}{2N_m}    \sum_{i,j,\mathbf R}
    b_i b_j
    \left[
    \frac{\mathbf I}{r_{ij,\mathbf R}^{3}}
    -
    3
    \frac{
    	\mathbf r_{ij,\mathbf R}
    	\mathbf r_{ij,\mathbf R}^{T}
    }{r_{ij,\mathbf R}^{5}}
    \right],
    $$
 
   Because of the long-range nature of the dipolar interactions, an accurate evaluation of $\mathbf{K}$ requires summation over the infinite lattice.
   We numerically evaluated its value and examined its convergence by progressively increasing the number of unit cells included in the summation.
   As shown in Fig.~\ref{fig:DipolarSum}(a), the diagonal elements converge to $K_{xx}=K_{yy}\approx-0.0180503$~\si{\per\angstromunit\cubed} and $K_{zz}\approx0.0361006$~\si{\per\angstromunit\cubed}, while the off-diagonal terms become negligible.
   
   Assuming a diagonal $g$-tensor, the dipolar energy per \Erion ion in temperature units can be written as
   $$
   E_{\rm dip}(\theta) = C_{\rm dip} S^2
   \left[
   g_{ab}^2 K_{xx}\sin^2\theta  +
   g_c^2 K_{zz}\cos^2\theta
   \right].
   $$
   where $C_{\rm dip}=\mu_0\mu_{\rm B}^{2}/(4\pi k_{\rm B})$, and $\theta$ denotes the angle between the spin direction and the $c$ axis.
   Because $K_{zz}>0$ and $K_{xx}<0$ for this A-type AFM structure, $E_{\rm dip}(\theta)$ is minimized  at $\theta=90^\circ$,
   regardless of the specific value of $g_{ab}$ and $g_c$ [Fig.~\ref{fig:DipolarSum}(b)].
   Thus, for the experimentally determined stacking pattern,
   long-range dipolar interactions favor an in-plane moment orientation, providing a plausible explanation for the contrast between the easy-axis anisotropy and the predominantly in-plane ordered moments.
   Furthermore, the energy difference between the $c$-axis ($\theta=0$) and $ab$-plane ($\theta=90^\circ$) moment orientations is further amplified by the extent of easy-axis anisotropy through
   $$
   \Delta E = E(0) - E(90^\circ) = C_{\rm dip} S^2 g_{ab}^2 (\alpha^2 K_{zz} - K_{xx}),
   $$
   where $\alpha = g_c/g_{ab}$ is the $g$-factor anisotropy ratio.
   Therefore, stronger easy-axis anisotropy further enhances the dipolar energetic preference for the in-plane moment orientation [Fig.~\ref{fig:DipolarSum}(b)].   

   To summarize, our results establish NaBaEr(BO$_3$)$_2$ as a rare-earth TLAF in which distinct sources of magnetic anisotropy  interplay on an exceptionally small energy scale.
   The clarified $R\bar{3}$ structure and CEF excitations establish the local environment and low-energy degrees of freedom of Er$^{3+}$, while the predominantly in-plane ordered moments contrast sharply with the pronounced easy-axis anisotropy of the ground-state doublet.
   Our dipolar-energy calculations show that long-range dipolar interactions favor the experimentally observed in-plane moment orientation for the A-type AFM structure.
   This result demonstrates that, on the small magnetic energy scale of NaBaEr(BO$_3$)$_2$, dipolar interactions can compete with CEF anisotropy in selecting the ordered-moment orientation.
   \NBEB therefore provides an intriguing platform for exploring the interplay between CEF effects and long-range dipolar interactions in rare-earth frustrated magnets.

	\begin{acknowledgments}
	This work was supported by the National Key Projects for Research and Development of China (Grant No. 2023YFA1406003), the National Natural Science Foundation of China (Grant No. 12574154, 12074023), the Fundamental Research Funds for the Central Universities in China, and the Large Scientific Facility Open Subject of Songshan Lake (Dongguan, Guangdong, No. KFKT2022B05). We acknowledge the support from the Synergetic Extreme Condition User Facility (SECUF, https://cstr.cn/31123.02.SECUF). ANSTO and ILL are gratefully acknowledged for providing neutron beam time through proposal No. P15836 and 4-05-919. We thank the staff members of the Multi-Physics Instrument (https://cstr.cn/31113.02.CSNS.MPI) at the China Spallation Neutron Source (CSNS) (https://cstr.cn/31113.02.CSNS ), for providing technical support and assistance in data collection and analysis.	
	\end{acknowledgments}
	\bibliography{NBEB-Ref.bib}

%
\clearpage
\onecolumngrid

\setcounter{figure}{0}
\renewcommand{\thefigure}{S\arabic{figure}}
\renewcommand{\theHfigure}{S\arabic{figure}}

\newcounter{smnote}
\newcommand{\smnote}[1]{\stepcounter{smnote}\section*{Note \thesmnote: #1}}

\begin{center}
	{\large Supplementary Materials for}
	
	$\,$\\
	\textbf{\large{Contrasting Crystal-Field and Dipolar Anisotropies in the Triangular Antiferromagnet \NBEB}}
	
	$\,$\\
	Su \textit{et al.}
\end{center}

\smnote{Estimation of the $c$-axis moment \wj{component}}

    \begin{figure}[h]
	   \centering
	   \includegraphics[width=0.9\textwidth]{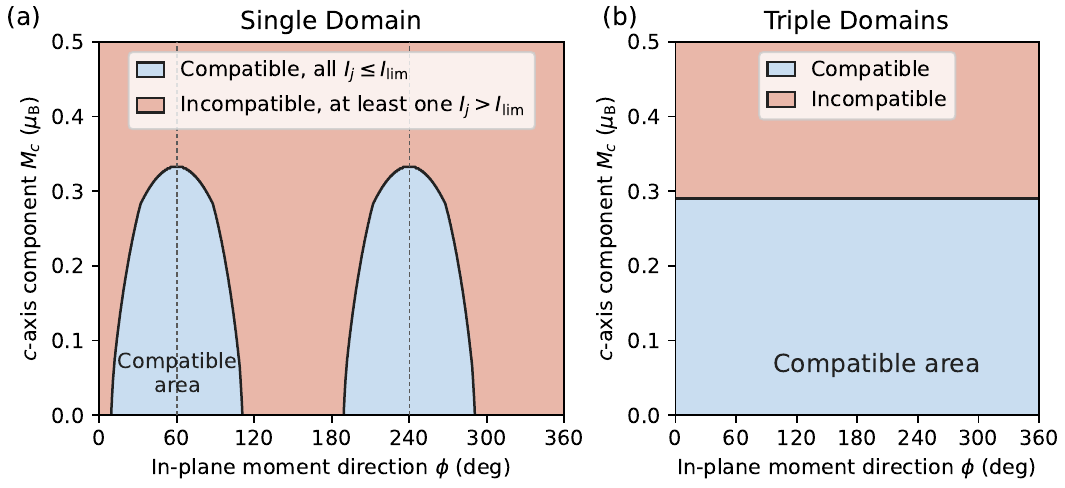}
	   \caption{
       Combinations of the in-plane moment direction ($\phi$) and the $c$-axis component ($M_c$) that are compatible (blue) or incompatible (red) with the absence of detectable magnetic reflections for the single-domain (a) and triple-domain (b) models.
       In both cases, $M_c$ is estimated to be less than $0.35~\mu_{\rm B}$. }
	\label{figSP:moment-c}
    \end{figure}
    
    A $c$-axis magnetic moment in \NBEB is symmetry allowed
    but not resolved in our diffraction experiment, presumably due to its
    small magnitude.
    Such a component would contribute to the 
    $(1,1,3n+1.5), (-1,-1,3n+1.5)$, and $(-2,-2,1.5)$ reflections, for which no magnetic 
    intensity was detected.
    Among the observed magnetic reflections, the weakest one is $(0,0,10.5)$, with an integrated intensity of 33(5).
    As a conservative estimate, we take two-thirds of this value, $I\rm_{lim}=$ 22,
    as the detection threshold under the present experimental signal-to-noise ratio.
    In other words, any $c$-axis component should not generate intensities exceeding this threshold, which allows us to estimate the upper bound on the $c$-axis component $M_c$.

    For simplicity, we assume that both \Erion sites share the same moment direction,
    $$
    \vec{M_1} = \vec{M_2} = (M_{ab} cos\phi, M_{ab}sin\phi, M_c),
    $$
    where $M_{ab} = 0.9$~\muB, $\phi$ denotes the in-plane angle with the $a$ axis,
    and $M_c$ is the $c$-axis component.
    We calculated the magnetic intensities $I_j$ for 
    $(1,1,3n+1.5)$, $(-1,-1,3n+1.5)$, and $(-2,-2, 1.5)$ reflections with
    n = 0--2, 4--5 as functions of $\phi \in (0,360^\circ)$ and $M_c \in (0,1)~\mu_{\rm B}$.
    Here reflections with n = 3 are excluded, because their $Q$ value is close to the powder-ring reflection from the sample environment.

    To satisfy the criterion of $I_j < I_{lim} = 22$ for all these reflections, $M_c$ has to be smaller than 0.35~\muB, as shown in the blue region in  Fig.~\ref{figSP:moment-c}.
    For the single-domain model, the allowed region favors an in-plane moment direction near (1, 1, 0)($\phi = 60 ^\circ$) [Fig.~\ref{figSP:moment-c}(a)].
    Including three symmetry-related orientation domains 
    removes this directional preference but leaves the upper bound on $M_c$ essentially unchanged [Fig.~\ref{figSP:moment-c}(b)].

\end{document}